\documentclass{article}
\usepackage[T1]{fontenc}

\makeatletter

\newcommand{\LyX}{L\kern-.1667em\lower.25em\hbox{Y}\kern-.125emX\spacefactor1000}

\csname @addtoreset\endcsname{equation}{section}
\newcommand{\dd}{\stackrel{\leftrightarrow}\partial}
\newcommand{\ddd}{\stackrel{\leftrightarrow}d}
\makeatother

\begin{document}

\begin{titlepage} 

\begin{flushright} {\small DFTT-15/99} \\ hep-th/9903086 \end{flushright} 

\vfill 

\begin{center} {

{\large On the Quantization of the GS Type IIB Superstring Action on \( AdS_{3}\otimes S_{3} \) with ~\( NSNS \) flux} 

\vskip 0.3cm

{\large {\sl }} 

\vskip 10.mm 

{\bf I. Pesando }\footnote{e-mail: ipesando@to.infn.it, pesando@alf.nbi.dk\\

Work supported by the European Commission TMR programme ERBFMRX-CT96-004}

\vskip4mm

{\small Dipartimento di Fisica Teorica , Universit\'a di Torino, via P. Giuria 1, I-10125 Torino, Istituto Nazionale di Fisica Nucleare (INFN) - sezione di Torino, Italy} 

}

\end{center} 

\vfill

\begin{center}{ \bf ABSTRACT}\end{center} 

\begin{quote}

We quantize the type IIB GS string action on \( AdS_{3}\otimes S_{3} \) with pure NSNS background flux and we show that it is equivalent to Liouville with periodic boundary condition plus free spacetime fermions.


\vfill

\end{quote} 

\end{titlepage}

\section{Introduction.}

The AdS/CFT conjecture (\cite{Maldacena},\cite{GubserKlenanovPolyako},\cite{Witten})
has in the last year attracted a lot of interest and work, one of its consequences
is that the classical (i.e. no string loops) type IIB superstring propagating
in a \( AdS_{5}\otimes S_{5} \) background is the masterfield of the \( {\cal N}=4 \)
\( D=4 \) SYM based on the \( U(N) \) gauge group. It has therefore become
important to write the classical action for the above mentioned type IIB superstring
and then to first quantize it. The first step has been accomplished. It started
with a series of works (\cite{MatsaevTseytlin},\cite{Kallosh},\cite{Kall-fixing})
which developed the now called supercoset approach, nevertheless the first proposal
for a superstring action which appeared in chronological order (\cite{Mio_ads})
was obtained using a different approach with respect to the above mentioned
supercoset approach, namely the supersolvable approach (\cite{gruppoToOsp})
while the second one (\cite{Kall_ads}) was obtained using the supercoset approach.
These two proposals were shown to be equivalent in (\cite{Mio_all}). Even if
the resulting action can be very easily written down and we can thrust it is
the proper conformal action (\cite{MatsaevTseytlin},\cite{kall-raj-vuoti}),
it is still unclear how to proceed to quantize it explicitly. Despite a recent
proposal (\cite{RajRoz}) on how to completely fix the conformal symmetry in
a perturbative way we are still far from a satisfactory way to quantize this
action, the main reason is that it describes a RR background which reflects
into the fact that the theory is not a usual WZW model because it has not the
left/right symmetry (\cite{MatsaevTseytlin}). In view of the fact that explicit
computations of \( {\cal N}=4 \) \( D=4 \) SYM amplitudes in the leading order
(\cite{4-punti}) and next to leading order (\cite{4-punti-corretta}) in \( \lambda  \)
( \( \lambda =Ng^{2}_{YM}>>1 \) ) based on this conjecture have revealed the
possible appearance of logarithm and even if the results in this field are far
from be definitive because of the difficulties of computing all the contributions
explicitly, it has become even more important to be able to understand how to
quantize this action.

Another unrelated aspect which is also very important is to gain a better understanding
of the holographic way of storing the dof on the boundary field theory (\cite{olografia}).

In order to get some hint on these problems we could start to look at some simpler
examples where one can at least partially compare with a well defined NSR formulation:
it is in fact purpose of this paper to quantize the explicit \( \kappa  \)
gauge fixed action for type IIB string propagating on \( AdS_{3} \) with a
pure NSNS background flux. This action was derived in (\cite{MIo_AdS3}), related
work on RR background can be found in (\cite{ParkRey}). The NSR formulation
has been worked out in (\cite{ads3-sei}) and further elaborated in (\cite{ooguri}),
see also (\cite{Yu}) for some work on light cone quantization and (\cite{BerVafWit})
for a new approach based on superalgebra WZW models. 

The main result of this paper is that all the gauge symmetries can be fixed
and the Virasoro constraints solved so that we are left with Liouville theory
with worldsheet periodic boundary conditions plus spacetime free fermions.

The plan of the paper is the following. In section 2 we write the action we
want to deal with. In section 3 we write the general solution of its equations
of motion. In section 3 we derive the Dirac brackets associated with our action.
In section 4 we discuss the gauge fixing of the residual conformal symmetry
after the choice of the conformal gauge and we solve the Virasoro constraints.
In section 5 we write the action to be quantized and we quantize it. Finally
in section 6 we discuss the meaning and the prospectives of this result.

\section{The GS action for pure NSNS background.}

We start writing the \( \kappa  \) fixed action for a type IIB superstring
propagating on \( AdS_{3}\times S_{3}\times T^{4} \) with pure NSNS background
using the notations of (\cite{MIo_AdS3}):

\begin{eqnarray}
S=-\frac{T}{2}\int _{\Sigma _{2}}d^{2}\xi \, \sqrt{-g}\, g^{\alpha \beta }\frac{1}{2}\times  &  & \nonumber \\
\left\{ 2e^{2}\: r^{2}\: \left[ \partial _{\alpha }x^{+}+\frac{i}{\sqrt{2}}\theta ^{(+)*n}\dd _{\alpha }\theta _{(+)n}\right] \right. \left[ \partial _{\beta }x^{-}+\frac{i}{\sqrt{2}}\theta ^{(-)*m}\dd _{\beta }\theta _{(-)m}\right]  &  & \nonumber \\
-\frac{1}{e^{2}}\frac{\partial _{\alpha }r\: \partial _{\beta }r}{r^{2}}-\frac{4}{e^{2}}\delta _{ij}\left. \frac{\partial _{\alpha }y^{i}\: \partial _{\beta }y^{j}}{(1-y^{2})^{2}}-\delta _{rs\: }\partial _{\alpha }z^{r}\: \partial _{\beta }z^{s}\right\}  &  & \nonumber \\
+|e|\: e\: r^{2}\left[ dx^{+}+\frac{i}{\sqrt{2}}\theta ^{(+)*n}\ddd \theta _{(+)n}\right] \wedge \left[ dx^{-}+\frac{i}{\sqrt{2}}\theta ^{(-)*m}\ddd \theta _{(-)m}\right]  &  & \nonumber \\
-\frac{1}{3}\rho e\: \int _{B_{3}}\frac{8}{|e|^{3}}\epsilon _{ijk}\frac{dy^{i}\wedge dy^{j}\wedge dy^{k}}{\left( 1-y^{2}\right) ^{3}} & \label{mia_azione} 
\end{eqnarray}
where \( x^{\pm } \) and \( r \) are the \( AdS_{3} \) coordinates, \( y^{i} \)
are the \( S_{3} \) projective coordinates, \( z^{r} \) are the \( T^{4} \)
coordinates and \( \theta ^{(\pm )}_{n} \) (\( n=1,2 \) ) are the surviving
fermionic coordinates after the fixing of the \( \kappa  \) gauge symmetry:
more precisely \( \theta _{(+)n} \) (\( \theta _{(-)n} \)) is the lower (upper)
component of a 3D spinor \( \Theta _{(+)n} \) (\( \Theta _{(-)n} \)). Moreover
\( \rho =\pm 1 \) is a free parameter and \( \dd =\overrightarrow{\partial }-\overleftarrow{\partial } \)

As derived (\cite{MIo_AdS3}) this action can only describe the states which
satisfy the condition
\begin{equation}
\label{stati_che_si_descrivono}
r^{2}\left( E^{-}_{0}-\frac{e}{\left| e\right| }E^{-}_{1}\right) \left( E^{+}_{0}+\frac{e}{\left| e\right| }E^{+}_{1}\right) \neq 0
\end{equation}
where \( E^{(\pm )}_{\alpha } \) is the component along \( d\xi ^{\alpha } \)
of the pullback of \( E^{(\pm )}=dx^{\pm }+\frac{i}{\sqrt{2}}\theta ^{(\pm )*n}\ddd \theta _{(\pm )n} \)
on the string worldsheet.

The first thing to notice is that this action (\ref{mia_azione}) can be written
as the sum of three independent terms. The first term is the WZW action for
the \( SU(2) \) group written in projective coordinates \( y^{i} \) when \( SU(2) \)
is reinterpreted as \( S_{3} \), the second term is the free action for \( T^{4} \)
in flat coordinates \( z^{r} \) and the third and interesting part is the action
for a superstring on \( AdS_{3} \). When we use \( d\xi ^{\alpha }\wedge d\xi ^{\beta }=\epsilon ^{\alpha \beta }d^{2}\xi  \)
the latter can be written as 
\begin{eqnarray*}
S_{AdS_{3}}=-\frac{T}{2}\int _{\Sigma _{2}}d^{2}\xi \, \sqrt{-g}\, \times \left\{ -\frac{1}{2e^{2}}g^{\alpha \beta }\frac{\partial _{\alpha }r\: \partial _{\beta }r}{r^{2}}\right.  &  & \\
|e|^{2}\left( g^{\alpha \beta }+\frac{e}{|e|}\frac{\epsilon ^{\alpha \beta }}{\sqrt{-g}}\right) \: r^{2}\: \left[ \partial _{\alpha }x^{+}+\frac{i}{\sqrt{2}}\theta ^{(+)*n}\dd _{\alpha }\theta _{(+)n}\right]  & \left. \left[ \partial _{\beta }x^{-}+\frac{i}{\sqrt{2}}\theta ^{(-)*m}\dd _{\beta }\theta _{(-)m}\right] \right\}  & \\
 & 
\end{eqnarray*}
If we choose \( e=-|e| \), we fix the conformal gauge \( g_{\alpha \beta }=e^{\phi }\eta _{\alpha \beta } \)
and we change coordinate as \( r=e^{\rho } \), the previous action becomes
equivalent to
\begin{eqnarray}
S_{AdS_{3}}=-\frac{T}{2}\int _{\Sigma _{2}}d^{2}\xi \,  & \left\{ -\frac{1}{e^{2}}\partial _{+}\rho \partial _{-}\rho \right.  & \nonumber \\
+2e^{2}\: e^{2\rho }\: \left[ \partial _{-}x^{+}+\frac{i}{\sqrt{2}}\theta ^{(+)*n}\dd _{-}\theta _{(+)n}\right]  & \left. \left[ \partial _{+}x^{-}+\frac{i}{\sqrt{2}}\theta ^{(-)*n}\dd _{+}\theta _{(-)n}\right] \right\}  & \nonumber \\
 & \label{basic_action} 
\end{eqnarray}
along with the Virasoro constraints
\begin{equation}
\label{virasoro}
2e^{2}\: e^{2\rho }\: \left[ \partial _{\pm }x^{+}+\frac{i}{\sqrt{2}}\theta ^{(+)*n}\dd _{\pm }\theta _{(+)n}\right] \left[ \partial _{\pm }x^{-}+\frac{i}{\sqrt{2}}\theta ^{(-)*n}\dd _{\pm }\theta _{(-)n}\right] -\frac{1}{e^{2}}\left( \partial _{\pm }\rho \right) ^{2}=0
\end{equation}
With the same choices the constraint on the states which can be described by
this action reads
\begin{equation}
\label{stati_buoni}
e^{2\rho }\left[ \partial _{-}x^{+}+\frac{i}{\sqrt{2}}\theta ^{(+)*n}\dd _{-}\theta _{(+)n}\right] \left[ \partial _{+}x^{-}+\frac{i}{\sqrt{2}}\theta ^{(-)*n}\dd _{+}\theta _{(-)n}\right] \neq 0
\end{equation}

\section{Equations of motion and their solution.}

It is now an easy matter to derive the bosonic equations of motion for this
pure \( NSNS \) background
\begin{eqnarray}
\partial _{+}\left( e^{2\rho }\: \left[ \partial _{-}x^{+}+\frac{i}{\sqrt{2}}\theta ^{(+)*n}\dd _{-}\theta _{(+)n}\right] \right)  & = & 0\nonumber \\
 &  & \label{eq_moto_x-} \\
\partial _{-}\left( e^{2\rho }\: \left[ \partial _{+}x^{-}+\frac{i}{\sqrt{2}}\theta ^{(-)*n}\dd _{+}\theta _{(-)n}\right] \right)  & = & 0\nonumber \\
 &  & \label{eq_moto_x+} \\
\partial _{+}\partial _{-}\rho +2e^{4}e^{2\rho }\left[ \partial _{-}x^{+}+\frac{i}{\sqrt{2}}\theta ^{(+)*n}\dd _{-}\theta _{(+)n}\right] \left[ \partial _{+}x^{-}+\frac{i}{\sqrt{2}}\theta ^{(-)*n}\dd _{+}\theta _{(-)n}\right]  & = & 0\nonumber \\
 &  & \label{eq_moto_rho} 
\end{eqnarray}
as well as the fermionic ones, of which we write just one because all the others
are very similar

\begin{equation}
\label{eq_moto_th+star}
\partial _{-}\theta _{(+)n}\: e^{2\rho }\: \left[ \partial _{+}x^{-}+\frac{i}{\sqrt{2}}\theta ^{(-)*n}\dd _{+}\theta _{(-)n}\right] =0
\end{equation}
We notice that to derive this equation we have made explicit use of eq. (\ref{eq_moto_x+}).
The important point to realize is that because of the constraint on the states
which can be described (\ref{stati_buoni}) this equation implies the free equation
of motion
\begin{equation}
\label{eq_moto_th1}
\partial _{-}\theta _{(+)n}=0
\end{equation}
The remaining fermionic equations of motion imply that all the \( \theta  \)
are also almost free (almost because as we will see their Dirac brackets are
not the free ones before the fixing of the residual conformal symmetry), explicitly
we find the following equations
\begin{equation}
\label{eq_moto_th2}
\partial _{-}\theta ^{*n}_{(+)}=\partial _{+}\theta _{(-)n}=\partial _{+}\theta _{(-)}^{*n}=0
\end{equation}
which can be trivially solved as
\begin{eqnarray}
\theta _{(+)} & = & \theta _{(+)}(\xi ^{+})\label{sol_th+} \\
\theta _{(-)} & = & \theta _{(-)}(\xi ^{-})\label{sol_th-} 
\end{eqnarray}

The bosonic equations can be solved explicitly too (\cite{ooguri}) since they
imply a kind of generalized Liouville equation; the solution reads 
\begin{eqnarray}
x^{+} & = & a(\xi ^{+})+k_{(+)}e^{-2c(\xi ^{+})}\frac{\overline{b}(\xi ^{-})}{1+b(\xi ^{+})\overline{b}(\xi ^{-})}\label{sol_x+} \\
x^{-} & = & \overline{a}(\xi ^{-})+k_{(-)}e^{-2\overline{c}(\xi ^{-})}\frac{b(\xi ^{+})}{1+b(\xi ^{+})\overline{b}(\xi ^{-})}\label{sol_x-} \\
\rho  & = & \lg \left( 1+b(\xi ^{+})\overline{b}(\xi ^{-})\right) +c(\xi ^{+})+\overline{c}(\xi ^{-})\label{sol_rho} 
\end{eqnarray}
with \( 2e^{4}k_{(+)}k_{(-)}+1=0 \) . Notice that \( k_{(+)} \) has opposite
sign wrt \( k_{(-)} \) but their absolute sign is not important since it can
be change by changing the sign of both \( b \) and \( \overline{b} \). 

Since we are dealing with a fundamental string all the coordinates have to be
periodic in \( \sigma =\xi ^{1} \) with period \( \pi  \) nevertheless this
does not imply that \( b(\xi ^{+}) \) has to be expanded in integer powers
of \( e^{2i\xi ^{+}} \). 

It is nice to verify that this solution explicitly implies the energy tensor
conservation that in turn implies that the Virasoro constraints can be used
to derive \( a \) and \( \overline{a} \) as done later.

\section{Dirac Brackets.}

We want now to examine the Dirac brackets associated with the action (\ref{basic_action})
since they are useful later when we discuss the final action. To this purpose
we proceed in the standard way and we define the conjugate momenta as 
\begin{eqnarray*}
{\cal P} & = & \frac{\partial {\cal L}}{\partial \dot{\rho }}=\frac{T}{2e^{2}}\dot{\rho }\\
{\cal P}_{-} & = & \frac{\partial {\cal L}}{\partial \dot{x}^{-}}=-\frac{Te^{2}}{\sqrt{2}}e^{2\rho }\: \left[ \partial _{-}x^{+}+\frac{i}{\sqrt{2}}\theta ^{(+)*n}\dd _{-}\theta _{(+)n}\right] \\
{\cal P}_{+} & = & \frac{\partial {\cal L}}{\partial \dot{x}^{+}}=-\frac{Te^{2}}{\sqrt{2}}e^{2\rho }\: \left[ \partial _{+}x^{-}+\frac{i}{\sqrt{2}}\theta ^{(-)*n}\dd _{+}\theta _{(-)n}\right] \\
 &  & 
\end{eqnarray*}
and similarly for the fermionic conjugate momenta. Since as usual we get first
class constraints, we have to use classical Dirac brackets which turn out to
be
\begin{eqnarray*}
\left\{ \rho (\sigma ),\: {\cal P}(\sigma ')\right\} =\left\{ x^{-}(\sigma ),\: {\cal P}_{-}(\sigma ')\right\} =\left\{ x^{+}(\sigma ),\: {\cal P}_{+}(\sigma ')\right\} =\delta (\sigma -\sigma ') &  & \\
\left\{ \theta ^{*n}_{(+)}(\sigma ),\: \theta _{(+)m}(\sigma )\right\} =-\frac{i}{\sqrt{2}{\cal P}_{+}}\delta ^{n}_{m}\: \delta (\sigma -\sigma ') &  & \\
\left\{ \theta ^{*n}_{(-)}(\sigma ),\: \theta _{(-)m}(\sigma )\right\} =-\frac{i}{\sqrt{2}{\cal P}_{-}}\delta ^{n}_{m}\: \delta (\sigma -\sigma ') &  & 
\end{eqnarray*}
The fermionic Dirac brackets confirm once again the necessity of imposing the
condition (\ref{stati_buoni}) on the states which can be described.

We notice that the previous brackets do not exhaust all the non vanishing brackets;
there is, for example, the bracket
\[
\left\{ x^{+}(\sigma ),\: \theta _{(+)n}(\sigma )\right\} =-\frac{1}{\sqrt{2}{\cal P}_{-}}\theta _{(+)n}\: \delta (\sigma -\sigma ')\]
We do not write them since they do not play any role in the following.

\section{Fixing of the residual conformal symmetry.}

We are now ready to write our proposal on how to fix the residual conformal
gauge symmetry \( \xi ^{+}\rightarrow \overline{\xi }^{+}=\overline{\xi }^{+}(\xi ^{+}) \)
and \( \xi ^{-}\rightarrow \overline{\xi }^{-}=\overline{\xi }^{-}(\xi ^{-}) \).
From a simple inspection of eq.s (\ref{sol_th+},\ref{sol_th-}) and eq.s (\ref{sol_x+}-\ref{sol_rho})
it is easy to derive that
\begin{eqnarray}
{\cal P}_{+} & = & -\frac{Te^{2}}{\sqrt{2}}\left( e^{2c}\partial _{+}b\right) (\xi ^{+})\label{P+_bc} \\
{\cal P}_{-} & = & -\frac{Te^{2}}{\sqrt{2}}\left( e^{2\overline{c}}\partial _{-}\overline{b}\right) (\xi ^{-})\label{p-_bcbar} 
\end{eqnarray}
We can now fix all the remaining conformal symmetry by imposing
\begin{eqnarray}
{\cal P}_{+} & = & \frac{p_{+}}{\pi }\label{gau_fix+} \\
{\cal P}_{-} & = & \frac{p_{-}}{\pi }\label{gau_fix-} 
\end{eqnarray}
where \( p_{\pm } \) are constants. Let us now show how this can be done; we
concentrate on \( {\cal P}_{+} \) since everything can be repeated exactly
in the same way for \( {\cal P}_{-} \). First of all we notice that \( {\cal P}_{+} \)
is actually the component of the 1-form \( \omega ={\cal P}_{+}d\xi ^{+} \),
then because of periodicity in \( \sigma  \) we can expand \( \omega ={\cal P}_{+}d\xi ^{+}=\left( {\cal P}_{+0}+\sum _{n\neq 0}{\cal P}_{+n}e^{2in\xi ^{+}}\right) d\xi ^{+} \).
Supposing that \( {\cal P}_{+0}\neq 0 \) we can then rewrite the previous expression
as \( \omega ={\cal P}_{+0}d\left( \xi ^{+}+\frac{i}{2}\sum _{n\neq 0}\frac{1}{n}\frac{{\cal P}_{+n}}{{\cal P}_{+0}}e^{2in\xi ^{+}}\right) ={\cal P}_{+0}d\overline{\xi }^{+} \)where
we have introduced the new variable \( \overline{\xi }^{+} \) thus fixing the
residual conformal symmetry. The last point to prove is that \( {\cal P}_{+0}\neq 0 \),
this follows from the constraint on the states which can be described (\ref{stati_buoni})
which implies \( {\cal P}_{+}(\xi ^{+})\neq 0 \) and by noticing that if \( {\cal P}_{+0}=0 \)
then \( \int _{0}^{\pi }d\sigma  \) \( {\cal P}_{+}=0 \) which would imply
that \( {\cal P}_{+} \) changes sign in the integration interval and hence
it has at least one zero.

Using eq.s (\ref{gau_fix+},\ref{gau_fix-}) and eq.s (\ref{P+_bc},\ref{p-_bcbar})
we can derive \( c \) (\( \overline{c} \)) as a functional of \( b \) (\( \overline{b} \)),
explicitly
\begin{eqnarray*}
e^{-2c} & = & -\frac{\pi Te^{2}}{\sqrt{2}p_{+}}k_{(-)}\partial _{+}b\\
e^{-2\overline{c}} & = & -\frac{\pi Te^{2}}{\sqrt{2}p_{-}}k_{(+)}\partial _{-}\overline{b}
\end{eqnarray*}
Inserting these results in the solution for \( \rho  \) (\ref{sol_rho}) we
can rewrite it as 
\begin{equation}
\label{rho=Liouville}
\rho =-\frac{1}{2}\log \left[ \left( \frac{\pi Te^{2}}{\sqrt{2p_{+}p_{-}}}\right) ^{2}k_{(+)}k_{(-)}\frac{\partial _{+}b\partial _{-}\overline{b}}{\left( 1+b\overline{b}\right) ^{2}}\right] 
\end{equation}
which is the general solution for the Liouville problem. The same result can
be reached substituting eq.s (\ref{gau_fix+},\ref{gau_fix-}) in the \( \rho  \)
equation of motion.

With the help of the gauge fixing conditions (\ref{gau_fix+},\ref{gau_fix-})
we can now solve the Virasoro constraints (\ref{virasoro}) as 
\begin{eqnarray}
x^{+}(\xi ^{+},\xi ^{-}) & = & -\frac{\pi T}{2\sqrt{2}p_{+}e^{2}}\int d\xi ^{+}\left( \partial _{+}\rho \right) ^{2}-\frac{i}{\sqrt{2}}\int d\xi ^{+}\theta ^{(+)*n}\dd _{+}\theta _{(+)n}+x^{+}_{0}\nonumber \\
 &  & \label{x+_da_Vir} \\
x^{-}(\xi ^{+},\xi ^{-}) & = & -\frac{\pi T}{2\sqrt{2}p_{-}e^{2}}\int d\xi ^{-}\left( \partial _{-}\rho \right) ^{2}-\frac{i}{\sqrt{2}}\int d\xi ^{-}\theta ^{(-)*n}\dd _{-}\theta _{(-)n}+x^{-}_{0}\nonumber \\
 &  & \label{x-_da_Vir} 
\end{eqnarray}
where \( x^{\pm }_{0} \) are the two zero modes conjugate to \( p_{\pm } \).
It is worth pointing out that eq.s (\ref{x+_da_Vir},\ref{x-_da_Vir}) are (as
it should, as it follows from the conservation of the energy momentum tensor)
compatible with the general solution for \( x^{\pm } \) (\ref{sol_x+},\ref{sol_x-}),
in fact we can deduce the explicit expression for \( a \) and \( \overline{a} \)
as
\begin{eqnarray*}
a(\xi ^{+}) & = & -\frac{\pi T}{2\sqrt{2}p_{+}e^{2}}\int d\xi ^{+}\left( \frac{\partial _{+}^{2}b}{\partial _{+}b}\right) ^{2}-\frac{i}{\sqrt{2}}\int d\xi ^{+}\theta ^{(+)*n}\dd _{+}\theta _{(+)n}+x^{+}_{0}\\
\overline{a}(\xi ^{-}) & = & -\frac{\pi T}{2\sqrt{2}p_{-}e^{2}}\int d\xi ^{-}\left( \frac{\partial ^{2}_{-}\overline{b}}{\partial _{-}\overline{b}}\right) ^{2}-\frac{i}{\sqrt{2}}\int d\xi ^{-}\theta ^{(-)*n}\dd _{-}\theta _{(-)n}+x^{-}_{0}
\end{eqnarray*}

\section{The quantum action.}

Up to now we have found that the only independent dof come from \( \rho  \)
and \( \theta  \)s since \( x^{\pm } \) can be expressed through them. To
understand which action governs their evolution after having solved the Virasoro
constraints we can rewrite their equations of motion 
\begin{eqnarray*}
\partial _{+}\partial _{-}\rho +2\frac{2p_{+}p_{-}}{(\pi T)^{2}}e^{-2\rho } & = & 0\\
p^{+}\partial _{-}\theta _{(+)n}=p^{+}\partial _{-}\theta ^{*n}_{(+)} & = & 0\\
p^{-}\partial _{+}\theta _{(-)n}=p^{-}\partial _{+}\theta ^{*n}_{(-)} & = & 0
\end{eqnarray*}
and their Dirac brackets
\begin{eqnarray*}
\left\{ \rho (\sigma ),{\cal P}(\sigma ')\right\}  & = & \delta (\sigma -\sigma ')\\
\left\{ \theta ^{*n}_{(+)}(\sigma ),\: \theta _{(+)m}(\sigma )\right\}  & = & -\frac{i\pi }{\sqrt{2}p_{+}}\delta ^{n}_{m}\: \delta (\sigma -\sigma ')\\
\left\{ \theta ^{*n}_{(-)}(\sigma ),\: \theta _{(-)m}(\sigma )\right\}  & = & -\frac{i\pi }{\sqrt{2}p_{-}}\delta ^{n}_{m}\: \delta (\sigma -\sigma ')
\end{eqnarray*}
These last two Dirac brackets suggest to introduce new grassman variables \( S_{(\pm )} \)
as
\begin{eqnarray*}
S_{(+)n}=\sqrt{2\sqrt{2}p_{+}}\theta _{(+)n} & \:  & S_{(+)}^{*n}=\sqrt{2\sqrt{2}p_{+}}\theta ^{*n}_{(+)}\\
S_{(-)n}=\sqrt{2\sqrt{2}p_{-}}\theta _{(-)n} & \:  & S_{(-)}^{*n}=\sqrt{2\sqrt{2}p_{-}}\theta ^{*n}_{(-)}
\end{eqnarray*}
so that the fermionic Dirac brackets become the canonical ones. Moreover these
new grassman variables can be assembled into \( 2 \) D=3 complex spinors as
follows
\[
S_{n}=\left( \begin{array}{c}
S_{(-)n}\\
S_{(+)n}
\end{array}\right) \]
since \( \theta _{(+)n} \) (\( \theta _{(-)n} \)) is the lower (upper) component
of a 3D spinor \( \Theta _{(+)n} \) (\( \Theta _{(-)n} \)).

After these redefinitions all of these previous equations of motion and Dirac
brackets can be derived from the action
\begin{equation}
\label{eq_fondam}
S_{AdS_{3}}=\int d^{2}\xi \frac{T}{2e^{2}}\left( \partial _{+}\rho \partial _{-}\rho +2\frac{2p_{+}p_{-}}{(\pi T)^{2}}e^{-2\rho }\right) +\frac{i}{2\sqrt{2}\pi }\left( S^{(+)*n}\dd _{-}S_{(+)n}+S^{(-)*n}\dd _{+}S_{(-)n}\right) 
\end{equation}

We can proceed to quantize the theory; as far as the Liouville part is concerned
we can proceed as in (\cite{Thorn}) where the proper periodic boundary conditions
are treated (see also (\cite{Gervais_Neveu}) for related work with different
boundary conditions) while for the fermionic part we get the usual anticommutators

\begin{eqnarray}
\left\{ S^{\dagger n}_{(+)}(\sigma ),\: S_{(+)m}(\sigma )\right\} =2\pi \delta ^{n}_{m}\delta (\sigma -\sigma ') &  & \label{rel_anti_S+} \\
\left\{ S^{\dagger n}_{(-)}(\sigma ),\: S_{(-)m}(\sigma )\right\} =2\pi \delta ^{n}_{m}\delta (\sigma -\sigma ') &  & \label{rel_anti_S-} 
\end{eqnarray}
To these (anti)commutation relations we have to add the commutation relations
for the \( x^{\pm } \) zero modes
\begin{equation}
\label{rel_comm_xp}
\left[ x^{+},\: p_{-}\right] =\left[ x^{-},\: p_{+}\right] =i
\end{equation}

As far as the equations which resolve the Virasoro constraints (\ref{x+_da_Vir},\ref{x-_da_Vir})
they have to be understood as normal ordered with respect to the oscillators
which satisfy the canonical (anti)commutation relations, this is specially true
for the ones associated with Liouville theory.

The last point to clarify is to understand which operator makes the string to
evolve in time; even if this treatment closely resemble the usual lightcone
quantization of the GS superstring in 10D here we find a non trivial worldsheet
hamiltonian and a ``trivial'', i.e. constant \( {\cal P}^{\pm } \) , therefore
we deduce the the right operator to use is the worldsheet hamiltonian.

\section{Conclusions.}

We can summarize the previous result by saying that we have shown that the GS
action for type IIB superstring on \( AdS_{3} \) with \( NSNS \) background
is equivalent to a Liouville theory with worldsheet periodic boundary condition
plus free fermions.

Up to now we have considered only the \( AdS_{3} \) part but it is very easy
to include the \( S_{3} \) part; this simply amounts to solve the Virasoro
constraints of the complete action, in particular we find for \( x^{\pm } \)
\begin{eqnarray*}
x^{+} & = & x^{+}_{AdS_{3}}-\frac{\pi T}{2\sqrt{2}p_{+}e^{2}}\int d\xi ^{+}T_{++}^{SU(2)-WZW}\\
x^{-} & = & x^{-}_{AdS_{3}}-\frac{\pi T}{2\sqrt{2}p_{-}e^{2}}\int d\xi ^{-}T_{--}^{SU(2)-WZW}
\end{eqnarray*}

We have now a clear quantum formulation of our problem which should allow us
to tackle some problems. The first one is to construct the quantum isometry
generators; this is important because we expect this background to be exactly
conformal and possible anomalies should show up in this algebra exactly as in
the usual flat 10D GS action where the critical dimensions is recovered from
the request of having an explicit realization of the 10D \( {N}=2 \) superPoincare'
algebra; to this purpose we should extend and make explicit the results obtained
in (\cite{kgener}). The second one is to perform the computations of scattering
amplitudes in this background in order to be able to see the stringy corrections
to the supergravity amplitudes. We could also consider branes in this GS formalism
as already done in the NSR formalism (\cite{dbrane}).

Another important point is what lesson we can learn from this example, we think
that this example shows once more how important it is to be able to find the
classical general solution, in fact both the WZWN models and this model are
exactly soluble because we know the general solution; this means that we can
hope to solve the more important problem of the \( AdS_{5}\otimes S_{5} \)
background only if we are able to find the general solution of the associated
string equations. 

\appendix

\section{Conventions.\label{app_A}}

We will use the following conventions:

\begin{itemize}
\item Indices: \( a,b,\ldots \in \{0,1,2\} \), \( u,v,\ldots \in \{3,\ldots ,9\} \),
\( i,j,\ldots \in \{3,4,5\} \), \( r,s,\ldots \in \{6,\ldots ,9\} \); WS indices
\( \alpha ,\beta ,\ldots \in \left\{ 0,1\right\}  \)
\item Epsilon: \( \epsilon _{012}=\epsilon _{345}=1 \); WS \( \epsilon _{01}=1 \)
\item metric: \( \eta _{ab}=diag(+1,-1,-1) \) , \( \eta _{ij}=diag(-1,-1,-1) \)
, \( \eta _{rs}=diag(-1,-1,-1,-1) \); WS \( \eta _{\alpha \beta }=diag(1,-1) \)
\item lightcone coordinates for \( AdS_{3} \) : \( x^{\pm }=\frac{1}{\sqrt{2}}\left( x^{0}\pm x^{1}\right)  \),
\( \epsilon _{+-2}=-1 \) 
\item WS lightcone coordinates : \( \xi ^{\pm }=\frac{\xi ^{0}\pm \xi ^{1}}{\sqrt{2}} \),
\( \xi _{\pm }=\frac{\xi _{0}\pm \xi _{1}}{\sqrt{2}} \); \( \eta _{+-}=\eta ^{+-}=1 \);
\( \epsilon _{+-}=-\epsilon ^{+-}=-1 \)
\item WS derivatives: \( \dd =\overrightarrow{\partial }-\overleftarrow{\partial } \);
\( \ddd =\overrightarrow{d}-\overleftarrow{d} \); \( \partial _{\pm }=\frac{\partial _{0}\pm \partial _{1}}{\sqrt{2}} \)
\end{itemize}

\end{document}